\documentclass[12pt,twoside,english]{article}
\usepackage[T1]{fontenc}
\usepackage[latin1]{inputenc}
\usepackage{a4wide}
\usepackage{graphicx}
\usepackage{setspace}
\usepackage{amssymb}

\makeatletter

 \newcommand{\lyxaddress}[1]{
   \par {\raggedright #1 
   \vspace{1.4em}
   \noindent\par}
 }

\usepackage{babel}
\makeatother
\begin{document}

\title{Equivalences between spin models induced by defects}

\author{Z. Bajnok}

\maketitle

\lyxaddress{\begin{center}Theoretical Physics Research Group of the Hungarian
Academy of Sciences, H-1117 P\'azm\'any s. 1/A Budapest, Hungary\end{center}}

\begin{abstract}
The spectrum of integrable spin chains are shown to be independent
of the ordering of their spins. As an application we introduce defects
(local spin inhomogeneities in homogenous chains) in two-boundary
spin systems and, by changing their locations, we show the spectral
equivalence of different boundary conditions. In particular we relate
certain nondiagonal boundary conditions to diagonal ones. 
\end{abstract}

\section{Introduction}

There have been an increasing interest in the recent years in two-boundary
spin systems. This is due to theoretical and practical reasons. On
one hand Nepomechie \cite{Raf2,Raf3} and Cao et al. \cite{Cao} were
able to extend the diagonal Bethe Ansatz (BA) solution of the XXZ
spin chain \cite{Diag} to certain nondiagonal cases where a constraint
is satisfied between the parameters of the two boundary conditions
(BCs). On the other hand these models have interesting applications
in recent problems of statistical physics with open boundaries such
as the description of the asymmetric exclusion process or the raise
and peel models, see for example \cite{GierEss,PRmod}. There are
developments in similar open spin chains as well, see e.g. the result
on XYZ and XXX models \cite{XYZ1,XYZ2,XXX}, but the basic example
is the XXZ model, from which other interesting models like the lattice
sine-Gordon, or the lattice Liouville models can be derived \cite{Anas1}. 

In the algebraic Bethe Ansatz method a pseudo vacuum vector is needed
on which the action of the monodromy matrix is triangular. In the
XXX \cite{XXX}, XXZ \cite{Cao} and XYZ \cite{XYZ1} models this
requirement resulted in a constraint relating the parameters of the
two boundary conditions. The same constraint appeared also in the
functional relations approaches in \cite{Raf2,Raf3} for the XXZ,
while in \cite{XYZ2} for the XYZ cases. Since they show up also in
the Temperly-Lieb formulation of the XXZ spin chains \cite{GierPy,Alex1}
we conclude that they are intrinsically encoded in the system.

It was observed in \cite{Alex1} that certain nondiagonal BCs in the
two-boundary XXZ spin chain are equivalent to diagonal ones. This
was then extended to other spin chains in \cite{Alex2}. 

The aim of the paper is to derive these equivalences and show the
physical origin of the constraints. In doing so we analyze inhomogeneous
spin chains and show, both in the periodic and in the open case, that
the spectrum of the transfer matrix/Hamiltonian does not depend on
the actual order of the spins merely on the spin content of the chain.
We define a different spin in a homogenous chain as a defect, then
we demonstrate how the equivalences observed in quasi periodic XXX
spin chains \cite{XXXp} can be re-derived by moving defects. In the
open case the nondiagonal BC is represented as a diagonal one dressed
by a defect. The idea is to move the defect, by performing similarity
transformations on the transfer matrix, to the other boundary in order
to dress that one instead. In the XXX model dressing the generic nondiagonal
BC gives rise to triangular BC if the constraint is satisfied. In
the XXZ model dressing the quantum group invariant BC gives diagonal
BC. Dressing the nondiagonal BC gives upper triangular type BC, whenever
the constraint is satisfied. Satisfying two constraints between the
two nondiagonal BCs the spectrum can be described by diagonal BCs
on both ends.

The paper is organized as follows: We start by deriving the equivalence
in the periodic chain. Having summarized the notations we introduce
defects and show that they can be shifted without altering the spectrum
of the transfer matrix. As a consequence we derive correspondences
between various periodic spin chains, and pedagogically illustrate
their usage on the example of the XXX model. Turning to the boundary
problem we recall the notations, then show how nondiagonal BCs can
be described by dressing diagonal ones with defects. The change of
the location of the defect from one side to the other leads to equivalences
between different BCs. In order to demonstrate the method we analyze
the XXX and the XXZ models in some detail.

\section{Equivalences in periodic chains}

Lets summarize the notations in the periodic case following the paper
of Sklyanin \cite{Skly}. We take a solution $R:\mathbb{C}\mapsto End(V\otimes V)$
of the Yang-Baxter equation (YBE)\begin{eqnarray}
R_{12}(u_{12})R_{13}(u_{13})R_{23}(u_{23}) & = & R_{23}(u_{23})R_{13}(u_{13})R_{12}(u_{12})\label{eq:YBE}\end{eqnarray}
 (with $u_{ij}=u_{i}-u_{j}$) which is symmetric $\mathcal{P}_{12}R_{12}\mathcal{P}_{12}=R_{12}^{t_{1}t_{2}}=R_{12}$
and satisfies unitarity and crossing symmetry:\begin{equation}
R_{12}(u)R_{12}(-u)=\rho(u)\quad;\quad R_{12}^{t_{1}}(u)R_{12}^{t_{1}}(-u-2\eta)=\tilde{\rho}(u)\label{eq:Unit}\end{equation}
Here $\rho$ and $\tilde{\rho}$ are scalar factors, $\mathcal{P}_{12}$
permutes the factors in the tensor product, $V\otimes V$, of the
vector space $V$, and the index refers to the factor in which the
operators act, i.e. the YBE is an equation in $V\otimes V\otimes V$
. The transposition in the $i$-th tensor product factor is denoted
by $t_{i}$. Graphically the YBE is represented as

\begin{center}\includegraphics[%
  height=2cm,
  keepaspectratio]{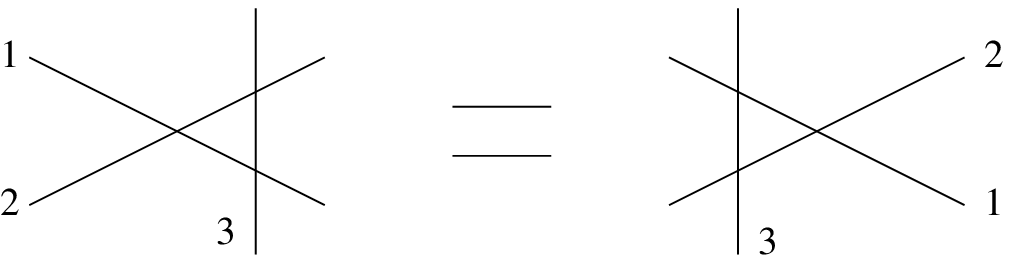}\end{center}

\noindent For such a solution of the YBE (\ref{eq:YBE}) we define
the associative algebra with generators $T_{ij}\quad i,j=1\dots dim\, V$
as\begin{equation}
R_{12}(u_{12})T_{1}(u_{1})T_{2}(u_{2})=T_{2}(u_{2})T_{1}(u_{1})R_{12}(u_{12})\label{eq:RTT}\end{equation}
Any representation can be used to define a spin chain, thus we deal
with concrete representations. If $T_{1}(u_{1})$ is represented in
the space $W_{a}$ then we denote it by $T_{1a}(u_{1})$, and the
equation takes the following form\begin{equation}
R_{12}(u_{12})T_{1a}(u_{1})T_{2a}(u_{2})=T_{2a}(u_{2})T_{1a}(u_{1})R_{12}(u_{12})\label{eq:RTTE}\end{equation}
which is called the RTT equation (RTTE) and can be written graphically
as

\begin{center}\includegraphics[%
  height=2cm,
  keepaspectratio]{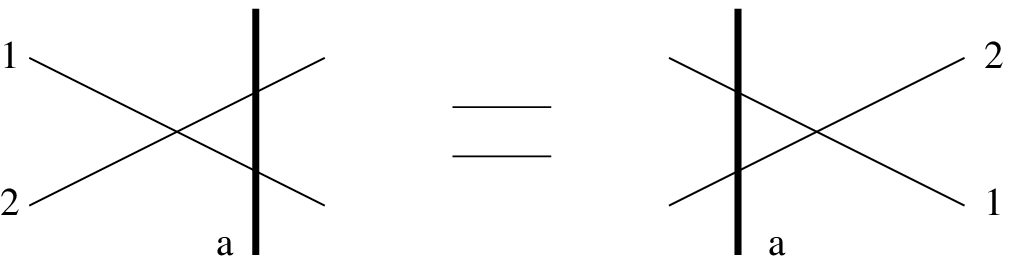}\end{center}

\noindent Whenever $T_{1a}$ and $T_{1b}$ are two representations
in $W_{a},W_{b}$, respectively, then $T_{1a}T_{1b}$ is also a representation
in $W_{a}\otimes W_{b}$. Moreover, if $T_{1a}(u)$ is a solution
of the RTTE (\ref{eq:RTTE}), then $T_{1a}^{-1}(-u)$ and $T_{1a}^{-1}(u)^{t_{1}}$
are also solutions of the same equation (\ref{eq:RTTE}).

Comparing the two figures (or the YBE (\ref{eq:YBE}) to the RTTE
(\ref{eq:RTTE})) we observe that we can always take $W_{a}=V$ and
choose $T_{ia}(u_{i})=R_{ia}(u_{i}+w)$, which, thanks to the YBE
(\ref{eq:YBE}), also satisfies the RTTE (\ref{eq:RTTE}).

In particular if the $R$-matrix, $R_{12}(u)$ is related to the universal
$R$-matrix, $\mathcal{R}$, of a quasi triangular Hopf algebra as
$R_{12}(u_{12})=(\pi_{u_{1}}\otimes\pi_{u_{2}})\mathcal{R}$ then
$(\pi_{u}\otimes I)\mathcal{R}$ provides a solution of the RTT algebra
relations (\ref{eq:RTT}) and any representation $\pi$ leads to a
concrete realization via $(\pi_{u}\otimes\pi)\mathcal{R}$. 

Suppose that $T_{1\pm}(u)$ are two solutions of the RTTE (\ref{eq:RTTE})
in the quantum spaces $W_{\pm}$, then the transfer matrices $t(u)=\textrm{Tr}{}_{1}(T_{1+}(u)T_{1-}(u))$
form a commuting family of matrices, i.e. they can be considered as
the generating functionals of conserved quantities for the quantum
system on $W_{+}\otimes W_{-}$. The typical example is as follows:
take a scalar representation on $W_{+}=\mathbb{C}$ with $T_{1+}(u)=K\in End(V)$
and another one on the quantum space $W_{-}=W\otimes\dots\otimes W$
of the form $T_{1-}(u)=L_{1a_{N}}(u)\dots L_{1a_{i}}(u)\dots L_{1a_{1}}(u)$
where the $L_{1a_{i}}$-s are the same representations. The transfer
matrix corresponds to a closed integrable quantum spin chain with
$N$ sites subject to quasi periodic boundary condition specified
by the matrix $K$, which can be represented graphically as:

\begin{center}\includegraphics[%
  height=21mm]{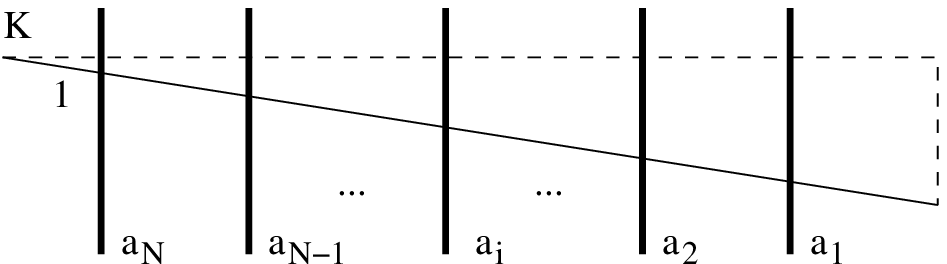}\end{center}

\noindent Note that $K=id$ is always a solution, which correspond
to periodic boundary condition. 

\noindent If at one position the representation is changed from $L_{1a_{i}}(u)$
to $L_{1a_{i}}^{'}(u)$ acting on the space $W^{'}$ instead of $W$
then we can interpret it as a \emph{defect} in the spin chain. We
can change all the representations to have a chain of the form $W_{-}=W_{N+1}\otimes\dots\otimes W_{i}\otimes\dots\otimes W_{1}$
with the corresponding solutions $L_{1a_{i}}^{i}$, where for $N+1$
we can take $a_{N+1}=\mathbb{C}$ and $L_{1a_{N+1}}^{N+1}=K$ and
choose $T_{1+}(u)=id$ to simplify the presentation. The transfer
matrix can be drawn as

\begin{center}\includegraphics[%
  height=22mm]{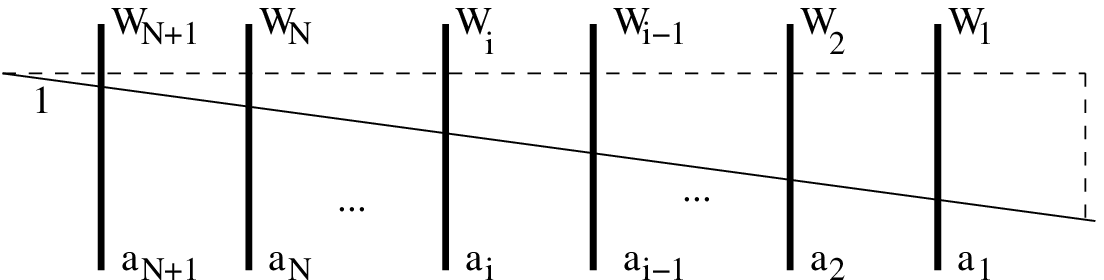}\end{center}

\noindent We claim that if there exist a matrix $S_{a_{i}a_{j}}(u_{ij})\in End(W_{i}\otimes W_{j})$
such that\begin{eqnarray}
S_{a_{i}a_{j}}(u_{ij})L_{1a_{i}}^{i}(u_{i})L_{1a_{j}}^{j}(u_{j}) & = & L_{1a_{j}}^{j}(u_{j})L_{1a_{i}}^{i}(u_{i})S_{a_{i}a_{j}}(u_{ij})\label{eq:SLL}\end{eqnarray}
which we call the SLL equation (SLLE) drawn graphically as

\begin{spacing}{0}
\begin{center}\includegraphics[%
  height=23mm,
  keepaspectratio]{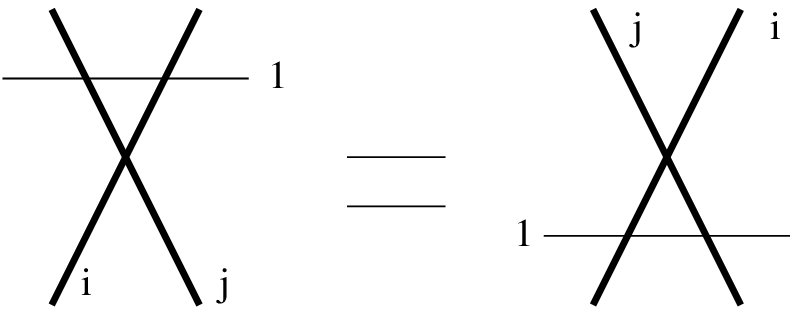}\end{center}
\medskip{}
\end{spacing}

\noindent then the spectrum of the transfer matrix does not depend
on the actual positions of the representations, merely on the representation
content of the chain. For this we show that we can change any two
neighboring operators. Take\begin{equation}
t(u)=\textrm{Tr}{}_{1}\bigl(L_{1a_{N+1}}^{N+1}(u)\dots L_{1a_{i+1}}^{i+1}(u)L_{1a_{i}}^{i}(u)L_{1a_{i-1}}^{i-1}(u)L_{1a_{i-2}}^{i-2}(u)\dots L_{1a_{1}}^{1}(u)\bigr)\label{eq:tper}\end{equation}
and use the SLLE (\ref{eq:SLL}) to replace $L_{1a_{i}}^{i}(u)L_{1a_{i-1}}^{i-1}(u)$
with $S_{a_{i}a_{i-1}}^{-1}(0)L_{1a_{i-1}}^{i-1}(u)L_{1a_{i}}^{i}(u)S_{a_{i}a_{i-1}}(0)$,
then commute the operator $S=S_{a_{i}a_{i-1}}(0)$ through the other
$L$-s since they act in different quantum spaces to show that\[
S\, t(u)\, S^{-1}=\textrm{Tr}{}_{1}\bigl(L_{1a_{N+1}}^{N+1}(u)\dots L_{1a_{i+1}}^{i+1}(u)L_{1a_{i-1}}^{i-1}(u)L_{1a_{i}}^{i}(u)L_{1a_{i-2}}^{i-2}(u)\dots L_{1a_{1}}^{1}(u)\bigr)\]
which can be drawn graphically as

\begin{center}\includegraphics[%
  height=22mm]{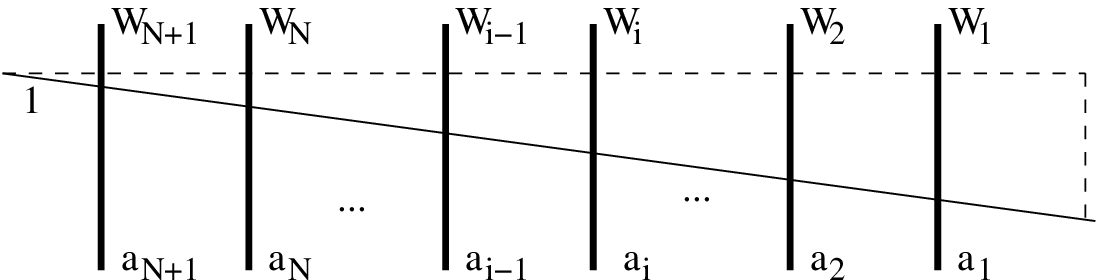}\end{center}

\noindent As a consequence if we have two defects then the actual
position of the defects does not matter, i.e., they do not interact.
Moreover, an alternating spin chain is equivalent to two equal length
homogenous chains coupled at two points together. Thus the results
on alternating spin chains \cite{Alter0,Alter1,Alter2,Alter3} can
be reinterpreted in this sense. 

The SLLE (\ref{eq:SLL}) looks very nontrivial, but it is almost always
satisfied in the spin chains considered sofar. As an example we mention,
that if $W_{i}=V$ and $W_{j}=W$ then the RTTE (\ref{eq:RTTE}) is
equivalent to the SLLE (\ref{eq:SLL}) by making the $R_{12}\to L_{12}$,
$T_{1a}\to L_{1a}$, and $T_{2a}^{-1}\to S_{2a}$ identification.
In general if a universal $\mathcal{R}$ matrix is given, then we
can always choose $S_{a_{i}a_{j}}(u_{ij})=(\pi_{u_{i}}^{a_{i}}\otimes\pi_{u_{j}}^{a_{j}})\mathcal{R}$
which provides a solution of the SLLE (\ref{eq:SLL}). Lets see some
concrete examples.

\subsection{Defects in the XXX spin chain }

Here for pedagogical reasons we reinterpret the results of \cite{XXXp}
for the $SU(2)$ invariant spin chain in our language. The R matrix
is taken to be \[
R_{12}(u)=\left(\begin{array}{cccc}
u+\eta & 0 & 0 & 0\\
0 & u & \eta & 0\\
0 & \eta & u & 0\\
0 & 0 & 0 & u+\eta\end{array}\right)\]
while for describing the spin $S_{j}$ at site $j$ we take the operator
\[
L_{1a_{j}}^{S_{j}}(u)=\left(\begin{array}{cc}
u+\frac{\eta}{2}+\eta S_{a_{j}}^{z} & \eta S_{a_{j}}^{-}\\
\eta S_{a_{j}}^{+} & u+\frac{\eta}{2}-\eta S_{a_{j}}^{z}\end{array}\right)\]
where $S_{a}$ represents $SU(2)$ with spin $S$. For $S=\frac{1}{2}$
we can recover the R-matrix itself $L_{1a}(u)=R_{1a}(u)$. Solving
the RTTE (\ref{eq:RTTE}) we can realize that we have also scalar
but $u$ independent solutions: \begin{equation}
T_{1}=\left(\begin{array}{cc}
A & B\\
C & D\end{array}\right)\quad;\qquad AD-BC=1\label{eq:xxxdef}\end{equation}
 and they correspond to the global $SU(2)$ symmetry of the model.
(Here we supposed invertibility of this matrix and normalized its
determinant to one). It can be used either as defects in the spin
chain or as the $L_{1N+1}=K$ matrix specifying quasi periodic boundary
conditions in (\ref{eq:tper}) as follows: Lets consider a chain depicted
on the next figure

\begin{center}\includegraphics[%
  height=30mm]{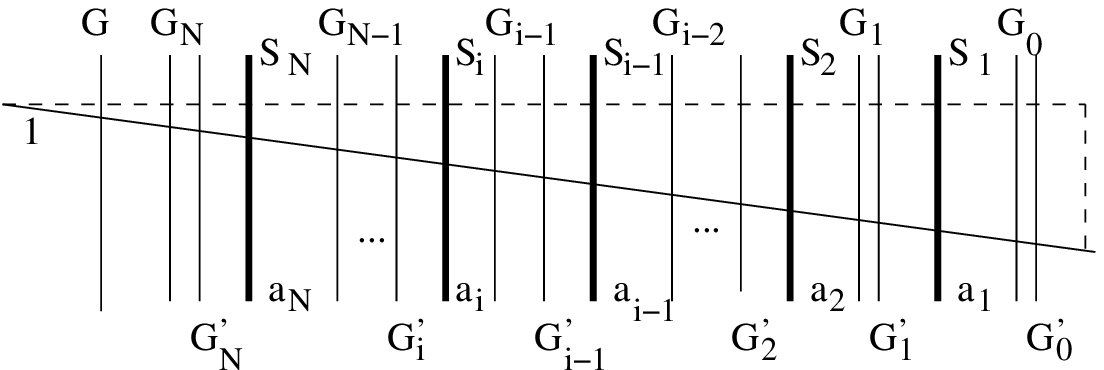}\end{center}

\noindent At site $i$ a spin $S_{i}$ representation is introduced
on the quantum space $a_{i}$ via the operator $L_{1a_{j}}^{S_{j}}$denoted
by a thick line. Between sites $i$ and $i-1$ two defects with matrices
of the form (\ref{eq:xxxdef}), denoted by $G_{i-1}$ and $G_{i-1}^{'}$,
are inserted. Finally quasi periodic boundary condition is introduced
by $G$ of the form (\ref{eq:xxxdef}). Thin lines correspond to scalar
solution (\ref{eq:xxxdef}) of the RTTE (\ref{eq:RTTE}). The $G_{i}^{'}L_{1a_{i}}^{S_{i}}G_{i-1}$
triple can be interpreted as the dressing of the spin $S_{i}$ with
defects and this transformation can be used to bring $L_{1a_{i}}^{S_{i}}$
to a triangular form. As was shown in Appendix B of \cite{XXXp} the
matrix $S$ exists in (\ref{eq:SLL}) so thick and thin lines can
be changed by similarity transformation. As a consequence we can move
the defects to the left and describe the spectrum by the following
transfer matrix:

\begin{center}\includegraphics[%
  height=25mm]{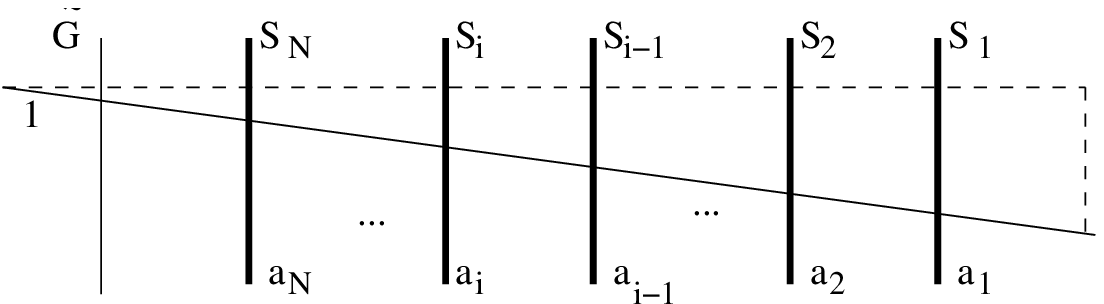}\end{center}

\noindent where $\tilde{G}=G\prod_{i=N}^{0}G_{i}G_{i}^{'}$ or any
of its cyclic permutation. Lets consider two applications. Take first
$G_{i}=id$ for $i>0$ and $G_{0}^{'}=G_{0}^{-1}$ and use cyclicity
to show that the spin chain with quasi periodic BCs specified by $G$
and $\tilde{G}=G_{0}^{-1}GG_{0}$ are equivalent. This transformation
can be used to diagonalize the matrix $G$. If we take now $G_{i}^{'}=id$
for all $i$ then we realize that we can 'collect' the defects into
$\tilde{G}=G\prod_{i}G_{i}$ and using the previous argument we conclude
that only the eigenvalues of the product of the defect matrices determine
the spectrum. All these statements can be checked on the explicit
BA solution of the model in \cite{XXXp}, where the independence of
the orders of the spins $S_{i}$ is also obvious. The generalization
of these results for the $SU(n)$ case following the lines in \cite{XXXp}
is straightforward.

\subsection{Defects in the XXZ spin chain}

Consider the XXZ spin chain, that is take $U_{q}(\hat{sl}_{2})$ to
be the quasi triangular Hopf algebra. The $R_{12}(u_{12})$ matrix
is the universal $\mathcal{R}$ matrix taken in the $\pi_{u_{1}}^{(1)}\otimes\pi_{u_{2}}^{(1)}$
representation:\begin{equation}
R_{12}(u)=\left(\begin{array}{cccc}
\sinh(u+\eta) & 0 & 0 & 0\\
0 & \sinh u & \sinh\eta & 0\\
0 & \sinh\eta & \sinh u & 0\\
0 & 0 & 0 & \sinh(u+\eta)\end{array}\right)\label{eq:RXXZ}\end{equation}
For $L_{1a}(u)$ we take the same representation written as\begin{equation}
L_{1a_{i}}(u)=\left(\begin{array}{cc}
\sinh(u+\frac{\eta}{2}(1+\sigma_{i}^{z})) & \sinh\eta\,\sigma_{i}^{-}\\
\sinh\eta\,\sigma_{i}^{+} & \sinh(u+\frac{\eta}{2}(1-\sigma_{i}^{z}))\end{array}\right)\label{eq:LXXZ}\end{equation}
where the operators $\sigma_{i}$ are the standard Pauli matrices.
The matrix structure always refers to the representation space, $\pi_{u_{1}}^{(1)}$
labeled by $1$. This defines a spin $\frac{1}{2}$ chain via the
transfer matrix (\ref{eq:tper}) as shown in \cite{Skly}. 

For the defect, which is relevant in the boundary problem, we take
the $\pi_{u}^{(1)}\otimes\pi_{u}^{q}$ representation, where $\pi_{u}^{q}$
is the q-oscillator ($q=e^{-\eta}$) representation of the Hopf algebra,
\cite{MarRob}. This gives the following defect operator\begin{equation}
T_{1a}(u,\beta,\mu_{1},\mu_{2})=\Gamma_{1}\left(\begin{array}{cc}
e^{u+\beta}q^{-J_{0}} & J_{-}q^{J_{0}}\\
-J_{+}q^{-J_{0}} & e^{u+\beta}q^{J_{0}}\end{array}\right)\Gamma_{2}^{-1}\label{eq:DFXXZ}\end{equation}
where $\Gamma_{i}=diag(e^{\mu_{i}/2},e^{-\mu_{i}/2})$ and the infinite
dimensional matrices, $J_{\pm},\, J_{0}$ can be written in terms
of the matrix units as\[
J_{0}=\sum_{j=-\infty}^{\infty}je_{jj}\quad;\quad J_{\pm}=\sum_{j=-\infty}^{\infty}e_{jj\mp1}\]
We have slightly changed the basis compared to \cite{MarRob}. The
matrices $\Gamma_{i}$ are the constant solutions of the RTTE (\ref{eq:RTTE})
and are related to the global symmetries of the model. They also can
be used to define quasi periodic boundary conditions as we have already
seen in the example of the XXX model. The analogous solution for the
defect equation in the sine-Gordon theory was analyzed in \cite{Andre,Ed1}.
It is clear from the previous considerations that if we have more
than one defect only their number, and not their locations, influences
the spectrum. The solution of the model even with one defect is an
open and interesting problem. Here we would like to concentrate on
how it can be used to derive equivalences in the open case and leave
this analysis for a future work.

\section{Equivalences in open spin chains}

In describing an open chain, additionally to the RTT algebra (\ref{eq:RTT}),
one also introduces two algebras spanned by $\mathcal{T}_{ij}^{\pm}$,
$i,j=1...dimV$. $\mathcal{T}^{-}$ corresponds to the right boundary
and satisfies\begin{eqnarray}
R_{12}(u_{1}-u_{2})\mathcal{T}_{1}^{-}(u_{1})R_{12}(u_{1}+u_{2})\mathcal{T}_{2}^{-}(u_{2})=\nonumber \\
 &  & \hspace{-5cm}\mathcal{T}_{2}^{-}(u_{2})R_{12}(u_{1}+u_{2})\mathcal{T}_{1}^{-}(u_{1})R_{12}(u_{1}-u_{2})\label{eq:BYBE}\end{eqnarray}
which is called the boundary YBE (BYBE) and is represented graphically
as

\begin{center}\includegraphics[%
  height=20mm]{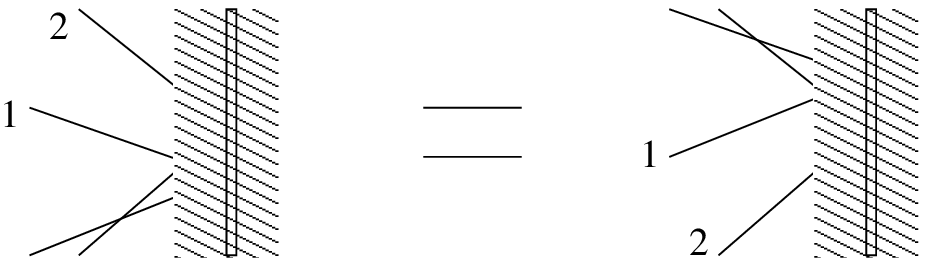}\end{center}

\noindent There is an analogous equation for $\mathcal{T}^{+}$ (cf.
\cite{Skly}). What we actually need is that the two algebras are
isomorphic: for any solution $\mathcal{T}^{-}(u)$ of the right BYBE
(\ref{eq:BYBE}) $\mathcal{T}^{+}(u)=\mathcal{T}^{-}(-u-\eta)^{t}$
solves the analogous left BYBE. In dealing with spin chains we are
interested in representations of these algebras $\mathcal{T}_{1a}^{\pm}$.
It is shown in \cite{Skly} that taken two solutions of the BYBE (\ref{eq:BYBE}),
$\mathcal{T}_{1}^{\pm}$, $t(u)=\textrm{Tr}{}_{1}(\mathcal{T}_{1}^{+}(u)\mathcal{T}_{1}^{-}(u))$
forms a commuting family of matrices and is the generating functional
for the integrals of motions of an open quantum system. It was also
shown in \cite{Skly} that taking two solutions $K_{1}^{\mp}(u)$
of the BYBE (\ref{eq:BYBE}) and a solution $T_{1a}$ of the RTTE
(\ref{eq:RTTE}) then $\mathcal{T}_{1a}^{-}(u)=T_{1a}(u)K_{1}^{-}(u)T_{1a}^{-1}(-u)$
solves the left (\ref{eq:BYBE}), while $\mathcal{T}_{1a}^{+}(u)=(T_{1a}(u)^{t_{1}}K_{1}^{+}(u)^{t_{1}}T_{1a}^{-1}(-u)^{t_{1}})^{t_{1}}$
the right BYBE. Choosing any of these \emph{dressed} solutions, $\mathcal{T}_{1a}^{\mp}$,
with the other undressed one, $K_{1}^{\pm}$, leads to the same transfer
matrix: \begin{equation}
t(u)=\textrm{Tr}{}_{1}(K_{1}^{+}(u)T_{1a}(u)K_{1}^{-}(u)T_{1a}^{-1}(-u))\label{eq:KTKT}\end{equation}
 Taking for $T_{1a}(u)$ the one used in the periodic chain $T_{1a}(u)=L_{1a_{N}}(u)\dots L_{1a_{i}}(u)\dots L_{1a_{1}}(u)$
the transfer matrix, $t(u)$, generates the conserved quantities for
an open spin chain with BCs specified by $K_{1}^{+}$ and $K_{1}^{-}$,
written graphically as: 

\begin{center}\includegraphics[%
  height=21mm]{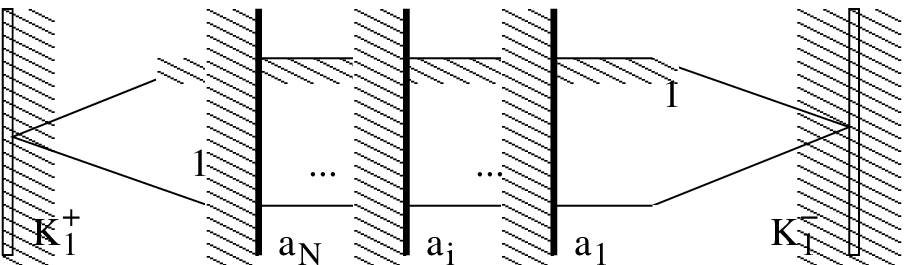}\end{center}

We can introduce defects by changing a representation at one site,
or, similarly to the periodic case, we can change the representations
on each site to have a more general chain with, \[
T_{1-}=L_{1a_{N}}^{N}(u)...L_{1a_{i+1}}^{i+1}(u)L_{1a_{i}}^{i}(u)L_{1a_{i-1}}^{i-1}(u)L_{1a_{i-2}}^{i-2}(u)...L_{1a_{1}}^{1}(u)\]
whose transfer matrix , $t(u)$, can be written graphically as

\begin{center}\includegraphics[%
  height=22mm]{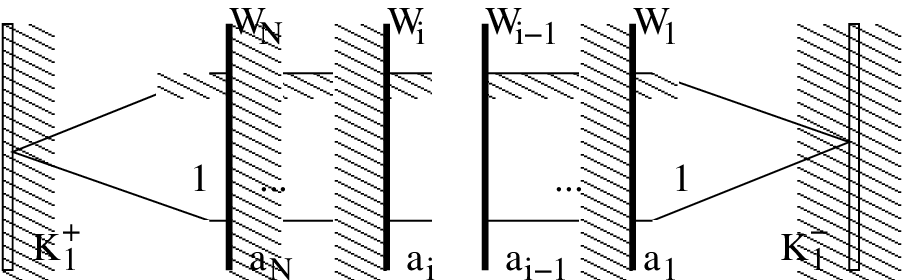}\end{center}

\noindent Now manipulations similar to those performed in the periodic
case show that any two neighboring representations can be changed.
This leads to a spectrally equivalent description by the transfer
matrix:

\begin{center}\includegraphics[%
  height=22mm]{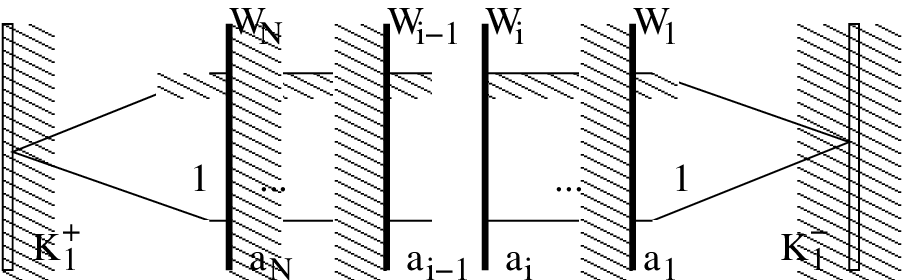}\end{center}

\noindent We conclude again that only the representation content matters
and not the actual order of the representations. In the typical applications
we take a chain with $N+1$ sites and interpret the first site with
operator $L_{1a_{1}}^{1}=T_{1a}(u)$ as a dressing of the boundary:
\begin{equation}
K_{1a}^{-}(u)=T_{1a}(u)K_{1}^{-}(u)T_{1a}^{-1}(-u)\label{eq:dr-}\end{equation}
 (For algebraic analysis of this type of dynamical BCs see \cite{DBC1,DBC2}.)
The defect can be moved to the other boundary to dress that one \begin{equation}
K_{1a}^{+}(u)^{t_{1}}=T_{1a}(u)^{t_{1}}K_{1}^{+}(u)^{t_{1}}T_{1a}^{-1}(-u)^{t_{1}}\label{eq:dr+}\end{equation}
 giving equivalences between different BCs, namely the system with
BCs $K_{1}^{-}$ and $K_{1a}^{+}$ is equivalent to the system with
different BCs described by $K_{1a}^{-}$ and $K_{1}^{+},$ moreover,
the equivalence is independent of the bulk spin content of the chain.
This isomorphism can map nondiagonal BCs to diagonal ones as we can
see in the next examples.

\subsection{Defects in the open XXX model}

Here we follow the presentation of the open XXX model of \cite{XXX}
but rewrite the results to our language. The simplest solution of
the BYBE (\ref{eq:BYBE}) has a diagonal form and contains one parameter\[
K_{1}^{-}(u,\bar{\xi}_{-})^{diag}=\left(\begin{array}{cc}
\bar{\xi}_{-}+u & 0\\
0 & \bar{\xi}_{-}-u\end{array}\right)\]
This solution can be dressed (\ref{eq:dr-}) by the defect $T_{1}$
in (\ref{eq:xxxdef}) to obtain the most general nondiagonal one\[
K_{1}^{-}(u,\xi_{-},c_{-},d_{-})=T_{1}K_{1}^{-}(u,\bar{\xi}_{-})^{diag}T_{1}^{-1}=\left(\begin{array}{cc}
\xi_{-}+u & c_{-}u\\
d_{-}u & \xi_{-}-u\end{array}\right)\]
where the parameters of the defect $T_{1}$ can be calculated from
$c_{-}$, $d_{-}$ and $\xi_{-}$ which is the dressed version of
$\bar{\xi}_{-}$. For the ratios we have two solutions with $\epsilon=\pm1$
as \[
\frac{A}{B}=\frac{1+\epsilon\sqrt{1+c_{-}d_{-}}}{c_{-}}\quad;\quad\frac{C}{D}=\frac{1-\epsilon\sqrt{1+c_{-}d_{-}}}{c_{-}}\]
Consider an open spin chain with bulk spins $S_{i}$ and boundary
conditions specified by $K_{1}^{-}(u,\xi_{-},c_{-},d_{-})$ on the
right and \[
K_{1}^{+}(u,\xi_{+},c_{+},d_{+})=\left(\begin{array}{cc}
\xi_{+}-u-\eta & -d_{+}(u+\eta)\\
-c_{+}(u+\eta) & \xi_{+}+u+\eta\end{array}\right)\]
on the left, which can be drawn graphically as

\begin{center}\includegraphics[%
  height=30mm,
  keepaspectratio]{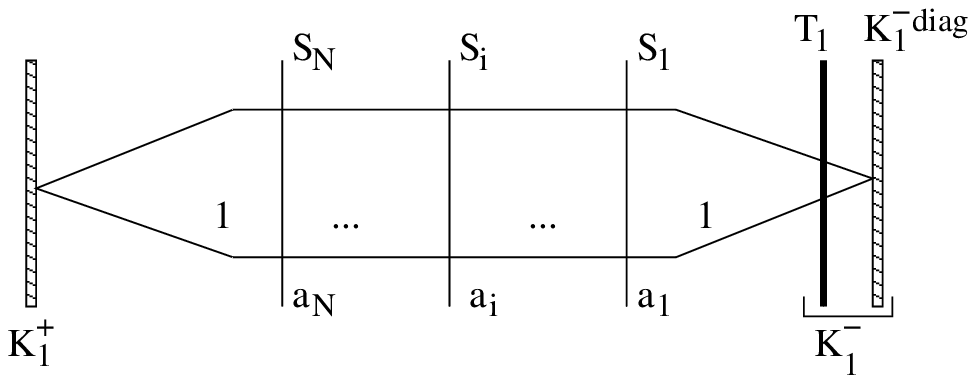}\end{center}

\noindent Just as in the bulk case we can move the defect to the other
boundary and then the transfer matrix is equivalently described as

\begin{center}\includegraphics[%
  height=30mm,
  keepaspectratio]{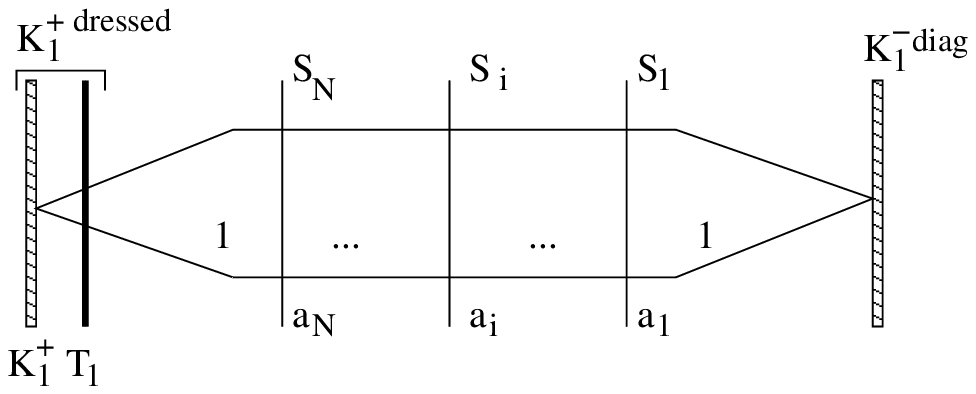}\end{center}

\noindent where now the left boundary is dressed (\ref{eq:dr+}) as
$K_{1}^{+dressed}=T_{1}^{-1}K_{1}^{+}T_{1}$. Demanding the upper/lower
triangularity of the matrix, which is needed to find a pseudo vacuum
in the BA formulation, we obtain the \[
\frac{1\pm\epsilon\sqrt{1+c_{-}d_{-}}}{c_{-}}=\frac{1+\epsilon\sqrt{1+c_{+}d_{+}}}{c_{+}}\]
 constraint. Under this condition BA equations for the eigenvalues
of the transfer matrix can be derived \cite{XXX}. If both conditions
are satisfied then the spin chain with two nondiagonal BCs is equivalent
to a chain with diagonal boundary conditions on both sides.

\subsection{Defects in the open XXZ model}

One of the simplest solutions of the BYBE (\ref{eq:BYBE}) is\begin{equation}
K_{1}^{-}(u)^{QGI}=diag(e^{u},e^{-u})\label{eq:QG}\end{equation}
which corresponds to the quantum group invariant (QGI) chain. By dressing
it (\ref{eq:dr-}) with the defect (\ref{eq:DFXXZ}) with parameters
$T_{1a}(u)=T_{1a}(u,\alpha_{-},\mu_{1},\mu_{2})$ and taking the $\mu_{1}=\to-\infty$
limit we can obtain the most general diagonal solution\begin{equation}
K_{1}^{-}(u,\alpha)^{diag}=\frac{n(u,\alpha)}{2}T_{1a}(u)K_{1}^{-QGI}T_{1a}^{-1}(-u)=\left(\begin{array}{cc}
P_{+} & 0\\
0 & P_{-}\end{array}\right)\label{eq:DIAG}\end{equation}
where $P_{\pm}=\cosh(u\pm\alpha_{-})$ and $n(u,\alpha)=e^{\alpha-2u}+e^{-\alpha}$.
Dressing it again (\ref{eq:dr-}) with the defect (\ref{eq:DFXXZ})\begin{equation}
T_{1a}(u)=T_{1a}(u,\beta_{-},\mu_{1}=\gamma_{-}-\alpha_{-},\mu_{2})\label{eq:TXXZ}\end{equation}
 we can obtain the following nondiagonal solution\begin{equation}
K_{1a}^{-}(u,\alpha_{-},\beta_{-},\gamma_{-})=n(u,\beta_{-})T_{1a}(u)K_{1}^{-}(u,\alpha)^{diag}T_{1a}^{-1}(-u)=\left(\begin{array}{cc}
P_{+}^{-} & J_{-}Q_{+}^{-}\\
J_{+}Q_{-}^{-} & P_{-}^{-}\end{array}\right)\label{eq:GENDYN}\end{equation}
where $P_{\pm}^{-}=e^{\beta_{-}}P_{\pm}+e^{-\beta_{-}}P_{\mp}$ and
$Q_{\pm}^{-}=\mp e^{\pm\gamma_{-}}\sinh(2u)$. The operators $J_{\pm}$
are the inverses of each other and can be diagonalized on the basis
$\vert\theta\rangle=\sum_{j=-\infty}^{\infty}e^{i\theta j}\vert j\rangle$
as $J_{\pm}\vert\theta\rangle=e^{\mp i\theta}\vert\theta\rangle.$
Thus each $\vert\theta\rangle$ subspace is invariant under the action
of $K_{1a}^{-}$ on which it takes the most general nondiagonal form

\begin{equation}
K_{1}^{-}(u,\alpha_{-},\beta_{-},\gamma_{-})=\left(\begin{array}{cc}
P_{+}^{-} & Q_{+}^{-}\\
Q_{-}^{-} & P_{-}^{-}\end{array}\right)\label{eq:GEN}\end{equation}
where the effect of the defect is the shift in the parameter $\gamma_{-}\to\gamma_{-}+i\theta$.
Our parameters are in spirit close to the parameterization used in
the sine-Gordon model \cite{Andre,Ed1}. They can be related to the
$(\hat{\alpha},\hat{\beta},\hat{\theta})$ parameters used in \cite{Raf3}
as $\hat{\alpha}=\alpha-\frac{i\pi}{2},\hat{\beta}=\beta,\hat{\theta}=\gamma+\frac{i\pi}{2}$. 

Lets consider the two-boundary spin $\frac{1}{2}$ XXZ chain with
nondiagonal BCs on the right end (\ref{eq:GEN}) specified by $K_{1a}^{-}(u,\alpha_{-},\beta_{-},\gamma_{-})$
and $K_{1}^{+}(u)$ on the left end. The transfer matrix can be represented
as

\begin{center}\includegraphics[%
  height=30mm,
  keepaspectratio]{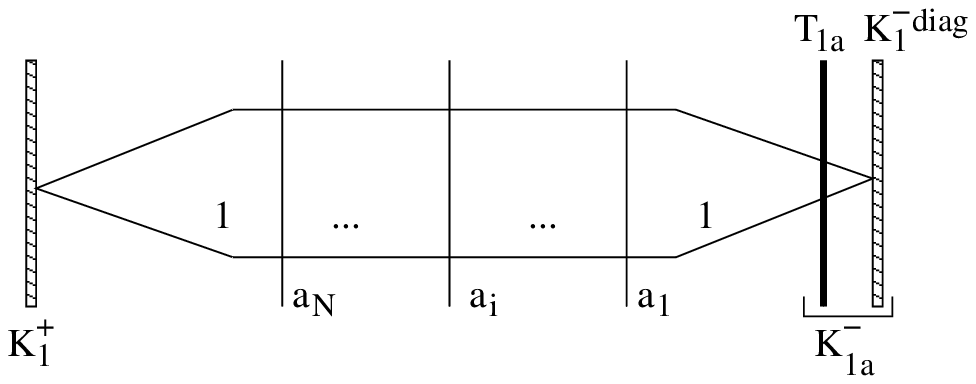}\end{center}

\noindent Moving the defect to the other boundary the transfer matrix
is equivalently described as

\begin{center}\includegraphics[%
  height=30mm,
  keepaspectratio]{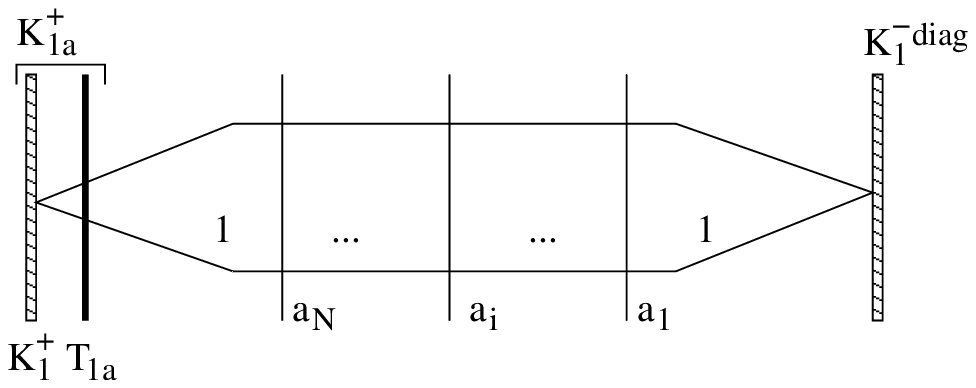}\end{center}

\begin{flushleft}where now the left boundary is dressed. \end{flushleft}

We start by deriving the equivalence found in \cite{Alex1}. In doing
so we take the QGI (\ref{eq:QG}) $K_{1}^{+}(u)=K_{1}^{-}(-u-\eta)^{QGI}$
BC on the left and the dressed diagonal $K_{1a}^{-}(u,\alpha_{-},\beta_{-},\gamma_{-})$
on the right (\ref{eq:GENDYN}). Although the defect depends on $\mu_{2}$
the dressed boundary and correspondingly the transfer matrix does
not. The dressed left boundary (\ref{eq:dr+}) takes the form \[
K_{1a}^{+}=\frac{1}{2}\left(\begin{array}{cc}
e^{\beta_{-}-u-\eta}+e^{-\beta_{-}+u+\eta} & e^{\mu_{2}}(e^{-2u-2\eta}-e^{2u+2\eta})q^{J_{0}}J_{-}q^{J_{0}}\\
0 & e^{\beta_{-}+u+\eta}+e^{-\beta_{-}-u-\eta}\end{array}\right)\]
Since the spectrum does not depend on $\mu_{2}$ we can take the limit
$\mu_{2}\to-\infty$ and now the dressed left boundary, $K_{1a}^{+}$
is diagonal $K_{1a}^{+}=K_{1}^{-}(-u-\eta)^{diag}$ with $\alpha_{-}\to\alpha_{+}=\beta_{-}$,
making equivalences between, nondiagonal BC $K_{1}^{-}(\alpha_{-},\beta_{-},\gamma_{-})$
on one end and QGI on the other, with diagonal BCs on both sides $K_{1}^{-}(\alpha_{-})^{diag}$
and $K_{1}^{+}(\beta_{-})^{diag}$ as was observed in \cite{Alex1}. 

As another application we take the most general nondiagonal $K_{1}^{+}(u)=K_{1}^{-}(-u-\eta)$
BC (\ref{eq:GEN}) with the $(\alpha_{+},\beta_{+},\gamma_{+})$ parameters:
$P_{\pm}^{+}(u)=P_{\pm}^{-}(-u-\eta)$ and $Q_{\pm}^{+}=Q_{\pm}^{-}(-u-\eta)$. 

The dressed transfer matrix takes the form \[
K_{1a}^{+}=\left(\begin{array}{cc}
P_{+a}^{+} & q^{-2J_{0}}e^{-\mu_{-}-\eta}Q_{+a}^{+}\\
q^{2J_{0}}e^{\mu_{-}-\eta}Q_{-a}^{+} & P_{-a}^{+}\end{array}\right)\]
\begin{eqnarray*}
P_{\pm a}^{+} & = & \sum_{\epsilon=\pm}(e^{\pm\epsilon\beta_{-}}P_{\epsilon}^{+}\mp e^{\pm\epsilon(u\pm\mu_{-})}Q_{\epsilon}^{+}J_{-\epsilon})\\
Q_{\pm a}^{+} & = & \sum_{\epsilon=\pm}\epsilon(e^{\pm\epsilon(u+\eta)}P_{\epsilon}^{+}\pm e^{\pm\epsilon(\beta_{-}+\eta\pm\mu_{-})}Q_{\epsilon}^{+}J_{-\epsilon})J_{\pm}\end{eqnarray*}
We see that the $\vert\theta\rangle$ subspace is not invariant under
the action of $K_{1a}^{+}$ since $q^{\pm J_{0}}\vert\theta\rangle=\vert\theta+i\eta\rangle$.
( In most of the cases $\eta$ is purely imaginary). If we demand
a highest/lowest weight property of the vector $\vert\theta\rangle$,
or some other words, the lower/upper triangularity of the matrix we
obtain the following constraint\begin{equation}
\beta_{+}\mp(\alpha_{+}-\gamma_{+})=\beta_{-}\mp(\alpha_{-}-\gamma_{-}-i\theta)+\eta\label{eq:CSTR}\end{equation}
which, using the discrete symmetries of the model, is the analog of
Nepomechie's constraint \cite{Raf2,Raf3,YNZ}. This condition is sufficient
to find a reference state in the BA \cite{Cao,Anas1}. The two choices
of signs are related as $\beta_{\pm}\leftrightarrow-\beta_{\pm}$
and $\eta\leftrightarrow-\eta$. This transformation does not change
the Hamiltonian whose spectrum we are describing (see \cite{Raf3})
merely its realization in terms of the transfer matrix. Similarly
the same Hamiltonian can be described in terms of different Temperly-Lieb
algebras \cite{GNPR} and the two constraints correspond to exceptional
representations of the two algebras, respectively. If we demand both
constraints \[
\beta_{+}=\beta_{-}+\eta\quad;\qquad\alpha_{+}-\gamma_{+}=\alpha_{-}-\gamma_{-}-i\theta\]
 the $\vert\theta\rangle$ subspace becomes invariant on which the
dressed left boundary matrix (\ref{eq:dr+}) takes the form $K_{1a}^{+}=f(u)K_{1}^{-}(-u-\eta,\alpha_{+})^{diag}$
with $f(u)=4\cosh(u+\beta_{+})\cosh(u-\beta_{-})$. This shows that
the two-boundary spin chain with nondiagonal BCs specified by $K_{1}^{+}$
and $K_{1a}^{-}$ can be described by a spin chain with diagonal BCs
specified by $K_{1}^{-diag}$ and $K_{1a}^{+}$ on the two ends. One
can check that under these circumstances the nondiagonal BA equations
\cite{Raf3} are equivalent to the diagonal ones \cite{Raf2}. 

We note that the equivalences derived here are valid for any value
of $\eta$. For special values, however, the defect admits a finite
dimensional representation, which might lead to other equivalences
observed in \cite{GNPR}, or help to understand the derivation of
BA equations for special cases in \cite{MN1,MN2}.

\section{Conclusion}

To sum up we have established an equivalence between different spin
chains: we have shown that under quite general (integrable) circumstances
the spectrum of the chain depends only on its representation (spin)
content and not on the actual order of the representations. We have
successfully applied this equivalence to make correspondence between
different BCs. We demonstrated the machinery on the examples of the
XXX and XXZ spin $\frac{1}{2}$ chain, where we mapped the system
with special nondiagonal BCs to diagonal ones. This equivalence is
quite general, however, and can be extended to other two-boundary
spin systems, like to the higher representation of $SU(2)$ or higher
rank algebras \cite{XXX,XXXp,Alex2}, in order for making equivalences
between different BCs. An advantage of the method is, that it is necessary
to solve one model from each equivalence class only. 

Finally we note that the analogue of dressing the boundary with defects
was already analised in conformal field theories \cite{Gerard}, in
integrable quantum field theories \cite{Def} and in classical field
theories \cite{Peter}. The straightforward generalization of our
ideas to these theories leads to equivalences between various BCs
in their two-boundary versions.

\section*{Acknowledgments}

We thank G. Tak\'acs, R. Nepomechie, A. Nichols and A. Doikou for
the discussions and useful comments. This work was supported by APCTP,
EUCLID HPRN-CT-2002-00325, OTKA K60040, T043582 and TS044839.

\end{document}